\documentclass[conference]{IEEEtran}
\ifCLASSINFOpdf
\else
\fi
\usepackage[cmex10]{amsmath}
\usepackage{amssymb}
\usepackage{mathtools}
\usepackage{epsfig}
\usepackage{epstopdf}
\usepackage{subfigure}
\usepackage{booktabs}
\usepackage{multirow}
\usepackage{blindtext}
\usepackage{algorithm}
\usepackage{algorithmic}
\usepackage{color}
\usepackage{amsmath}
\usepackage{graphicx}
\usepackage[noadjust]{cite}
\usepackage{array}
\usepackage[hang,flushmargin]{footmisc}

\newcommand{\ab}{\mathbf{a}}
\newcommand{\Ab}{\mathbf{A}}
\newcommand{\Abt}{\widetilde{\Ab}}

\newcommand{\Ib}{\mathbf{I}}
\newcommand{\Pb}{\mathbf{P}}
\newcommand{\pb}{\mathbf{p}}
\newcommand{\Qb}{\mathbf{Q}}
\newcommand{\Rb}{\mathbf{R}}
\newcommand{\Lb}{\mathbf{L}}
\newcommand{\Lbh}{\widehat{\Lb}}
\newcommand{\Mb}{\mathbf{M}}
\newcommand{\Sb}{\mathbf{S}}
\newcommand{\sib}{\mathbf{s}}

\newcommand{\wb}{\mathbf{w}}
\newcommand{\Wb}{\mathbf{W}}
\newcommand{\Xb}{\mathbf{X}}
\newcommand{\xb}{\mathbf{x}}

\newcommand{\Zerb}{\mathbf{0}}

\newcommand{\Rbb}{\mathbb{R}}
\newcommand{\DimDef}{\Rbb^{n_1 \times n_2}}

\newcommand{\PCal}{\mathcal{P}_{\Omega^c}}

\newcommand{\thetb}{\boldsymbol{\theta}}

\newcommand{\TSNR}{\text{SNR}}
\newcommand{\tetb}{\boldsymbol{\theta}}
\newcommand{\Rw}{\Rb_{\wb}}
\newcommand{\Rwh}{\widehat{\Rb}_{\wb}}
\newcommand{\Rx}{\Rb_{\xb}}
\newcommand{\Rxh}{\widehat{\Rb}_{\xb}}
\newcommand{\Rs}{\Rb_{\sib}}
\newcommand{\Rst}{\widetilde{\Rb}_{\sib}}
\newcommand{\Rsha}{\widehat{\Rb}_{\sib}}

\DeclareMathOperator{\rank}{rank}

\DeclareMathOperator{\diag}{diag}
\DeclareMathOperator{\vect}{vec}

\DeclareMathOperator*{\argmin}{argmin}

\def\RzCol{black}

\IEEEoverridecommandlockouts

\begin{document}
%
\title{DOA Estimation in Partially Correlated Noise Using Low-Rank/Sparse Matrix Decomposition}

\author{\IEEEauthorblockN{Mohammadreza Malek-Mohammadi\IEEEauthorrefmark{1}\IEEEauthorrefmark{2}, Magnus Jansson\IEEEauthorrefmark{2}, Arash Owrang\IEEEauthorrefmark{2},\\Ali Koochakzadeh\IEEEauthorrefmark{1}, Massoud Babaie-Zadeh\IEEEauthorrefmark{1}}
\IEEEauthorblockA{\IEEEauthorrefmark{1}Sharif Univ.~of Tech., Tehran, Iran, Email: {\color{\RzCol}\{mrezamm, ali\_k\}@ee.sharif.edu}, mbzadeh@sharif.edu}
\IEEEauthorblockA{\IEEEauthorrefmark{2}{\color{\RzCol}ACCESS Linnaeus Centre,} KTH, Stockholm, Sweden, Emails: {\color{\RzCol}\{janssonm, owrang\}@kth.se}} \thanks{{\color{\RzCol}This work was supported by the Swedish Research Council under contract 621-2011-5847 and the Iran National Science Foundation under contract 91004600.}}
}
\maketitle

\begin{abstract}
We consider the problem of direction-of-arrival (DOA) estimation in unknown partially correlated noise environments where the noise covariance matrix is sparse. A sparse noise covariance matrix is a common model for a sparse array of sensors consisted of several widely separated subarrays. Since interelement spacing among sensors in a subarray is small, the noise in the subarray is in general spatially correlated, while, due to large distances between subarrays, the noise between them is uncorrelated. Consequently, the noise covariance matrix of such an array has a block diagonal structure which is indeed sparse. Moreover, in an ordinary nonsparse array, because of small distance between adjacent sensors, there is noise coupling between neighboring sensors, whereas one can assume that nonadjacent sensors have spatially uncorrelated noise which makes again the array noise covariance matrix sparse. Utilizing some recently available tools in low-rank/sparse matrix decomposition, matrix completion, and sparse representation, we propose a novel method which can resolve possibly correlated or even coherent sources in the aforementioned partly correlated noise. In particular, when the sources are uncorrelated, our approach involves solving a second-order cone programming (SOCP), and if they are correlated or coherent, one needs to solve a computationally harder convex program. We demonstrate the effectiveness of the proposed algorithm by numerical simulations and comparison to the Cramer-Rao bound (CRB).
\end{abstract}


%
\IEEEpeerreviewmaketitle

\section{Introduction}
The assumption of spatially white noise in an array of sensors (antennas) is violated in many practical scenarios. For example, when the antennas are closely spaced, the small interelement spacing leads to strong mutual coupling between array elements \cite{Bala12}. 
A consequence of this coupling would be correlation between the noise of array elements. It is known that the performance of conventional direction-of-arrival (DOA) estimation methods degrades significantly when the noise is spatially correlated (colored) \cite{LoV92,PesaG01,StoiS92}. Colored noise in an antenna array can also be present due to environmental conditions \cite{Wenz62}. 
Nevertheless, the problem of DOA estimation in an unknown spatially colored noise is not solvable without some restrictions on the impinging sources or on the noise field \cite{StoiS92}. A popular solution is to exploit some largely spaced subarrays in which due to large distance between these subarrays, the inter-subarray noise is uncorrelated. 
These configurations for sensor arrays are also known as sparse arrays.

Different algorithms have been proposed to use this type of arrays to estimate the DOA that are mainly based on the maximum likelihood (ML) criterion; see e.g., \cite{LiN11,VoroGW05}. 
However, ML approaches lead to solving some nonconvex optimization problems which are generally very hard to solve and there is no guarantee for convergence to the global optimum solution. Moreover, the ML approaches are only derived under the assumption of Gaussian data.

In this paper, we propose a new algorithm based on matrix rank minimization and sparse representation techniques which can effectively estimate the directions of possibly correlated emitters in environments where the noise covariance matrix is unknown but sparse by solving a convex optimization program. Particularly, this algorithm can be used when a sparse array is exploited, the noise field is nonuniform (the noise covariance matrix is diagonal but every diagonal entry is arbitrary) \cite{PesaG01}, or only there is noise coupling between adjacent sensors. Also, it is worth mentioning that we will not impose any assumption on the distribution of the noise and sources; we only assume that they are zero-mean and stationary random processes.

The rest of this paper is organized as follows. After formulating the problem in Section \ref{sec:Form}, we introduce our method in Section \ref{sec:Algo} and present some numerical examples in Section \ref{sec:Sim}. Section \ref{sec:Con} concludes the paper.


\section{Problem Formulation} \label{sec:Form}

Consider an array of $m$ antennas and assume that $q$ sources are impinging on this array. Further, assume that the propagation time of the received signals across the array is much less than the inverse of the signal bandwidth (the assumption of being narrow-band). Samples at the output of antennas can be formulated according to the model
\begin{equation} \label{maineq}
\xb(n) = \Ab(\tetb) \sib(n) + \wb(n), ~~ n = 1,\cdots,N,
\end{equation}
where $\xb(n) = \big(x_1(n),\cdots,x_m(n)\big)^T$ denotes the vector of samples at time instant $n$ from antenna 1 to $m$, $N$ is the total number of collected samples, $\Ab(\tetb)=\big[\ab(\theta_1),\cdots,\ab(\theta_q)]$ is the array manifold at unknown directions $\thetb = (\theta_1,\cdots,\theta_q)^T$, $\sib(n) = \big(s_1(n),\cdots,s_q(n)\big)^T$ designates the vector of source signals at time instant $n$, and $\wb(n) = \big(w_1(n),\cdots,w_m(n)\big)^T$ is the vector of noise at different antennas.


\section{The proposed approach} \label{sec:Algo}
First, we briefly review the concepts of matrix completion (MC) and low-rank/sparse matrix decomposition which are used in the derivation of our algorithm.
\subsection{Introduction}
In the matrix completion problem, we observe some entries of a matrix and want to recover other unobserved elements \cite{CandR09}. Generally, it is not possible to reconstruct a matrix from a subset of its entries. However, if the matrix is low-rank and the position of revealed entries follows a certain random law, then using
\begin{equation} \label{MCRank}
\min_{\Xb} \rank(\Xb)~\text{s.t.}~ [\Xb]_{ij} = [\Mb]_{ij},~ (i,j) \in \Omega,
\end{equation}
in which $\Mb \in \DimDef$ is the low-rank matrix to be reconstructed and $\Omega \subset \{1,\cdots,n_1\} \times \{1,\cdots,n_2\}$ is the index set of observed entries, one can recover $\Mb$ with high probability \cite{CandR09}. The convex relaxation of \eqref{MCRank} leads to
\begin{equation} \label{MCNuc}
\min_{\Xb} \|\Xb\|_*~\text{s.t.}~ [\Xb]_{ij} = [\Mb]_{ij},~ (i,j) \in \Omega,
\end{equation}
where $\| \Xb \|_* = \sum_{i=1}^{r} \sigma_i(\Xb)$ denotes the nuclear norm of matrix $\Xb$ in which $\sigma_i(\Xb)$ is the $i$th largest singular value of $\Xb$ and $r = \rank(\Xb)$. Under more restrictive conditions, solving \eqref{MCNuc} results in obtaining the unique solution of \eqref{MCRank} \cite{CandR09}.

When the observations are contaminated by additive noise, i.e., $\Xb = \Mb + \Wb$, where $\Wb$ is a matrix modelling the additive noise, \eqref{MCNuc} can be updated to
\begin{equation} \label{MCNucNoisy}
\min_{\Xb} \|\Xb\|_* + \lambda_{MC} \sum_{i,j \in \Omega} \big([\Xb]_{ij} - [\Mb]_{ij}\big)^2,
\end{equation}
where $\lambda_{MC} > 0$ is some constant to regularize between being low-rank and consistency with noisy observations.

Now, suppose that we have a matrix $\Xb \in \DimDef$ which is equal to the sum of a low-rank and a sparse matrix. More precisely,
\begin{equation*}
\Xb = \Lb + \Sb,
\end{equation*}
where $\Lb$ is a low-rank matrix and $\Sb$ is a sparse matrix in which only a few entries are nonzero. The problem of decomposing $\Xb$ into $\Lb$ and $\Sb$ is underdetermined in general since the number of unknowns is larger than the number of equations. This task can be formulated as
\begin{equation} \label{RPCARank}
\min_{\Lb,\Sb} \rank(\Lb) + \gamma_1 \| \Sb\|_0 ~\text{s.t.}~ \Xb = \Lb + \Sb,
\end{equation}
in which $\gamma_1 > 0$ is a regularization parameter and $\|\cdot\|_0$ denotes the number of nonzero entries of a matrix.

It has been shown that, under some mild assumptions, solving \eqref{RPCARank} recovers the matrices $\Lb$ and $\Sb$ \cite{CandLMW11}. Nonetheless, this problem is NP-hard. The tightest convex relaxation of \eqref{RPCARank} equals \cite{CandLMW11}
\begin{equation} \label{RPCAConv}
\min_{\Lb,\Sb} \| \Lb \|_* + \gamma_2 \| \Sb\|_1 ~\text{s.t.}~ \Xb = \Lb + \Sb,
\end{equation}
where {\color{\RzCol}$\| \Sb\|_1 = \sum_{i=1}^{n_1} \sum_{j=1}^{n_2} | [\Sb]_{ij} |$}.

Under some mild deterministic or probabilistic conditions, \eqref{RPCARank} and \eqref{RPCAConv} share the same unique solution \cite{
ChanSPW11,CandLMW11}.
When $\Xb = \Lb + \Sb + \Wb$, where $\Wb$ is a additive noise, \eqref{RPCAConv} is updated to
\begin{equation} \label{RPCAConv_Noisy}
\min_{\Lb,\Sb} \| \Lb \|_* + \gamma_D \| \Sb\|_1 +\lambda_D \| \Xb - \Lb - \Sb \|_F^2,
\end{equation}
where, similar to \eqref{MCNucNoisy}, $\lambda_D$ is some regularization parameter and $\|\cdot\|_F$ designates the Frobenius norm.

\subsection{The main idea}
The main idea of our approach to estimate the vector of unknown directions $\tetb$ relies on the decomposition of the sample covariance matrix. To be precise, assuming sources and noise are uncorrelated, from \eqref{maineq}, we have
\begin{equation} \label{Rxeq}
\Rx = \Ab \Rs \Ab^H + \Rw,
\end{equation}
where $\Rx = E\{\xb(n) \xb(n)^H\}$, $\Rs = E\{\sib(n) \sib(n)^H\}$, and $\Rw = E\{\wb(n) \wb(n)^H\}$ are covariance matrices.

It can be verified that $\rank(\Ab \Rs \Ab^H) \leq q$; thus, if the number of sources is much smaller than the number of antennas, then $\Ab \Rs \Ab^H$ will be a low-rank matrix. Furthermore, we assume that $\Rw$ is an unknown matrix but sparse. As discussed in Section I, this assumption can be satisfied in a sparse array of antennas or when there is noise coupling between adjacent sensors.\footnote{$\Rw = \sigma^2 \Ib$ and $\Rw = \diag(\sigma_1^2,\cdots,\sigma_m^2)$ are also sparse covariance matrices and can be handled by the proposed algorithm.}
For instance, when a uniform linear array (ULA) is exploited and the noise of neighboring sensors is correlated, $\Rw$ may have the following structure
\begin{equation} \label{RwStruct}
\Rw = \begin{bmatrix}
               \sigma_1^2     & \sigma_{1,2} & 0                & 0                & \cdots & 0\\
               \sigma_{2,1}   & \sigma_{2}^2 & \ddots           & 0                & \cdots & 0\\
               0              & \ddots       & \ddots           & \ddots           & 0      & \vdots\\
               \vdots         & 0            & \ddots           & \ddots           & \ddots & 0 \\
               0              & \cdots       & 0                & \sigma_{m-1,m-2} & \sigma_{m-1}^2 & \sigma_{m-1,m}\\
               0              & \cdots       & 0                & 0                & \sigma_{m,m-1} & \sigma_{m}^2
         \end{bmatrix}.
\end{equation}

In summary, to estimate DOAs, we make the following assumptions.
\begin{itemize}
  \item A1: The noise and sources are zero-mean wide-sense random processes and are uncorrelated.
  \item A2: The radiated sources can be correlated or even coherent.
  \item A3: The noise covariance matrix is arbitrary but sparse. The support of this matrix, location of nonzero entries, are known from, for example, the geometry of the array.
  \item A4: The number of sources is unknown and much smaller than the number of antennas.
\end{itemize}

As a first solution, we can exploit program \eqref{RPCAConv_Noisy} to recover $\Ab \Rs \Ab^H$ and $\Rw$ from the matrix $\Rx$. However, using the above assumptions more efficiently, we can exploit the information that we know the support of $\Rw$ to obtain better results. Let $\Omega$ denote the support set of $\Rw$ and $\PCal$ be a projection to the set $\Omega^c = \{1,\cdots,m\} \times \{1,\cdots,m\} \setminus \Omega$ such that
\begin{equation*}
\PCal(\Xb) = \left \{
    \begin{array}{ll}
    0 &  (i,j) \in \Omega,\\
	\protect[\Xb\protect]_{ij} & \text{otherwise},
	\end{array} \right.
\end{equation*}
Applying $\PCal$ on \eqref{Rxeq}, we get
\begin{equation*}
\PCal(\Rx) = \PCal(\Ab \Rs \Ab^H).
\end{equation*}
Consequently, the task of estimating $\Ab \Rs \Ab^H$ simplifies to a MC problem,
\begin{equation*}
\min_{\Xb} \| \Xb \|_*~\text{s.t.}~ \PCal(\Xb) = \PCal(\Rx).
\end{equation*}

However, in practice, only an estimate of $\Rx$ is available. Let
\begin{equation*}
\Rxh = \frac{1}{N} \sum_{n=1}^{N} \xb(n) \xb(n)^H
\end{equation*}
designate the sample covariance matrix, then we have $\Rxh = \Ab \Rs \Ab^H + \Rw + \Qb$, where $\Qb$ is the disturbance term due to finite number of samples. Particularly, when sources and noise have normal distributions, $\Qb$ has a recentered-Wishart distribution \cite{EatoE83}. To mitigate the effect of finite samples, we use the following program to recover $\Ab \Rs \Ab^H$
\begin{equation} \label{LEst}
\Lbh = \argmin_{\Xb} \{\| \Xb \|_*  + \lambda_1 \| \PCal(\Xb) - \PCal(\Rxh) \|_F ~ | ~\Xb \succeq \Zerb \},
\end{equation}
where $\Xb \succeq \Zerb$ means that $\Xb$ is a positive semidefinite matrix. {\color{\RzCol}In \eqref{LEst} and other optimization programs we use in what follows, the data fidelity terms (e.g., $\| \PCal(\Xb) - \PCal(\Rxh) \|_F$ in \eqref{LEst}) are not squared. This lets us to select the regularization parameter similar to \cite{BellCW11} independent from scaling the covariance of $\Qb$.} If the support of $\Rw$ is not known, one can use
\begin{multline*}
(\Lbh,\Rwh) = \argmin_{(\Lb,\Sb)} \{ \| \Lb \|_* + \gamma_D \| \Sb \|_1 \\
 + \lambda_D \| \Rxh - \Lb - \Sb \|_F ~ | ~\Lb \succeq \Zerb, \Sb \succeq \Zerb \}.
\end{multline*}
to estimate $\Ab \Rs \Ab^H$.

In the next step, we need to estimate $\tetb$ from $\Lbh$, an estimate of $\Ab \Rs \Ab^H$. As $\Ab$ is unknown, we use a gridding technique to find DOAs. Let $\Abt = [\ab(\phi_1), \cdots, \ab(\phi_M)]$ denote the sampled array manifold in which $\phi_1, \cdots, \phi_M$ are the grid directions and $M$ is the number of grid points. If the gridding is fine enough, then $\Ab \Rs \Ab^H \approx \Abt \Rst \Abt^H$, where $\Rst$ equals to $\Rs$ in rows and columns associated to $\phi_k \approx \theta_i,~ 1 \leq i \leq q$, and is zero in other locations.

As a result of this gridding, we use the following optimization problem to estimate $\Rst$
\begin{equation} \label{REst}
\Rsha = \argmin_{\Pb} \{\| \Pb \|_1  + \lambda_2 \|\Lbh - \Abt \Pb \Abt^H \|_F ~ | ~\Pb \succeq \Zerb \}.
\end{equation}
After obtaining $\Rsha$ from the above program, $\diag(\Rsha)$ designates the estimated spatial spectrum at the grid points.

Also, it is possible to combine \eqref{LEst} and \eqref{REst} to solve directly for $\Rsha$, i.e.,
\begin{multline} \label{REst2}
\Rsha = \argmin_{\Pb} \{\| \Abt \Pb \Abt^H \|_* + \alpha \| \Pb \|_1 \\
+ \beta \| \PCal(\Abt \Pb \Abt^H) - \PCal(\Rxh) \|_F ~ | ~\Pb \succeq \Zerb \}.
\end{multline}
However, because we have to choose two regularization parameters at the same time, solving \eqref{REst2} may be harder than estimating $\Rst$ in two steps. In contrast, when sources are uncorrelated, $\Pb$ is a diagonal matrix and with {\color{\RzCol}letting} $\pb = \diag(\Pb)$, \eqref{REst2} simplifies to
\begin{equation} \label{UncorrEst}
\min_{\pb} \| \pb \|_1 + \lambda_u \| \PCal((\Abt^* \odot \Abt) \pb) - \PCal( \vect(\Rxh) ) \|_2 ~ \text{s.t.} ~\pb \succeq \Zerb,
\end{equation}
where $\Abt^{*}$ denotes the conjugate of $\Abt$, $\odot$ is the Khatri-Rao product (column-wise Kronecker product), $\vect(\Rx)$ denotes the vector with the columns of $\Rx$ stacked on top of one another, $\|\cdot\|_2$ is the $\ell_2$-norm, and $\pb \succeq \Zerb$ means that all entries of $\pb$ are non-negative.\footnote{{\color{\RzCol}After submitting this paper, we became aware that a special case of \eqref{UncorrEst}, where $\Rw$ is diagonal, has been proposed in \cite{HeSH14}. However, \eqref{UncorrEst} applies to a more general setting and includes an appropriate choice for $\lambda_u$.}}

\section{Numerical simulations} \label{sec:Sim}
In this section, the performance of the proposed algorithm is numerically analyzed and is compared to the {\color{\RzCol}stochastic CRB which can be obtained by extending the stochastic CRB for nonuniform white noise in \cite{PesaG01}}. In the simulations, we use a 10-element ULA with half wavelength antenna spacing. Sources and sensors are at the same plane. Signals and noise are iid realizations of zero-mean Gaussian distributions with covariance matrices $\Rx$ and $\Rw$, respectively. Further, the noise covariance matrix in all experiments has the structure given in \eqref{RwStruct} with $\sigma_1^2,\cdots,\sigma_m^2$ equal to 1, $\sigma_{1,2}, \cdots \sigma_{m-1,m} = 0.5j$, and $\sigma_{2,1}, \cdots \sigma_{m,m-1} = -0.5j$.

\begin{figure}[tb]
\centering
\includegraphics[width=0.49\textwidth]{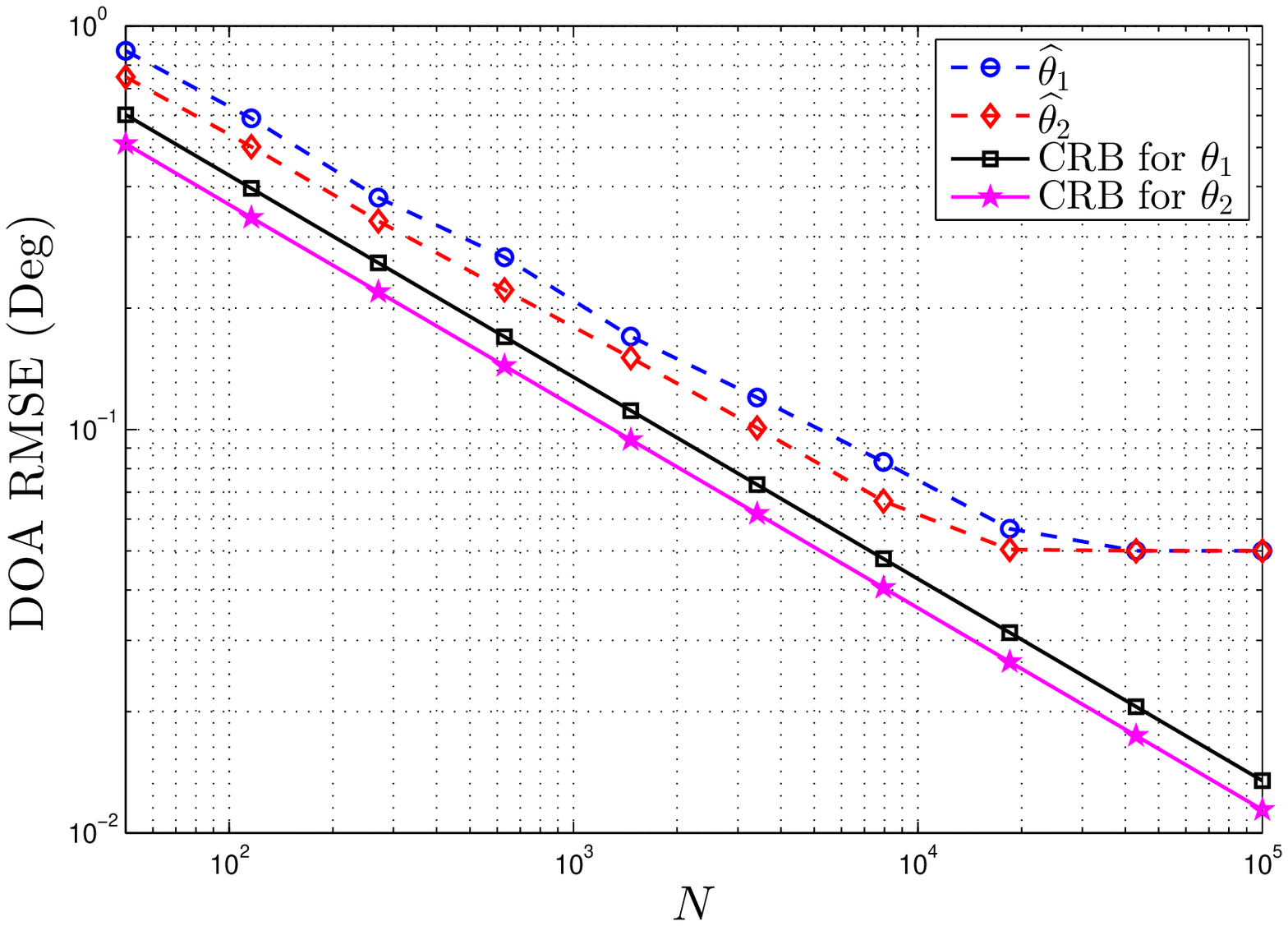}
\vspace{-0.75cm}
\caption{RMSEs for estimation of $\theta_1$ and $\theta_2$ using the proposed program \eqref{UncorrEst} as well as corresponding CRBs are plotted as a function of number of samples. True $\theta_1$ and $\theta_2$ are $88.05^{\circ}$ and $91.95^{\circ}$, respectively. 500 Monte-Carlo simulations are run and $\TSNR = 0 \text{ dB}$.} \label{fig:FigN}
\end{figure}

\begin{figure}[tb]
\centering
\includegraphics[width=0.49\textwidth]{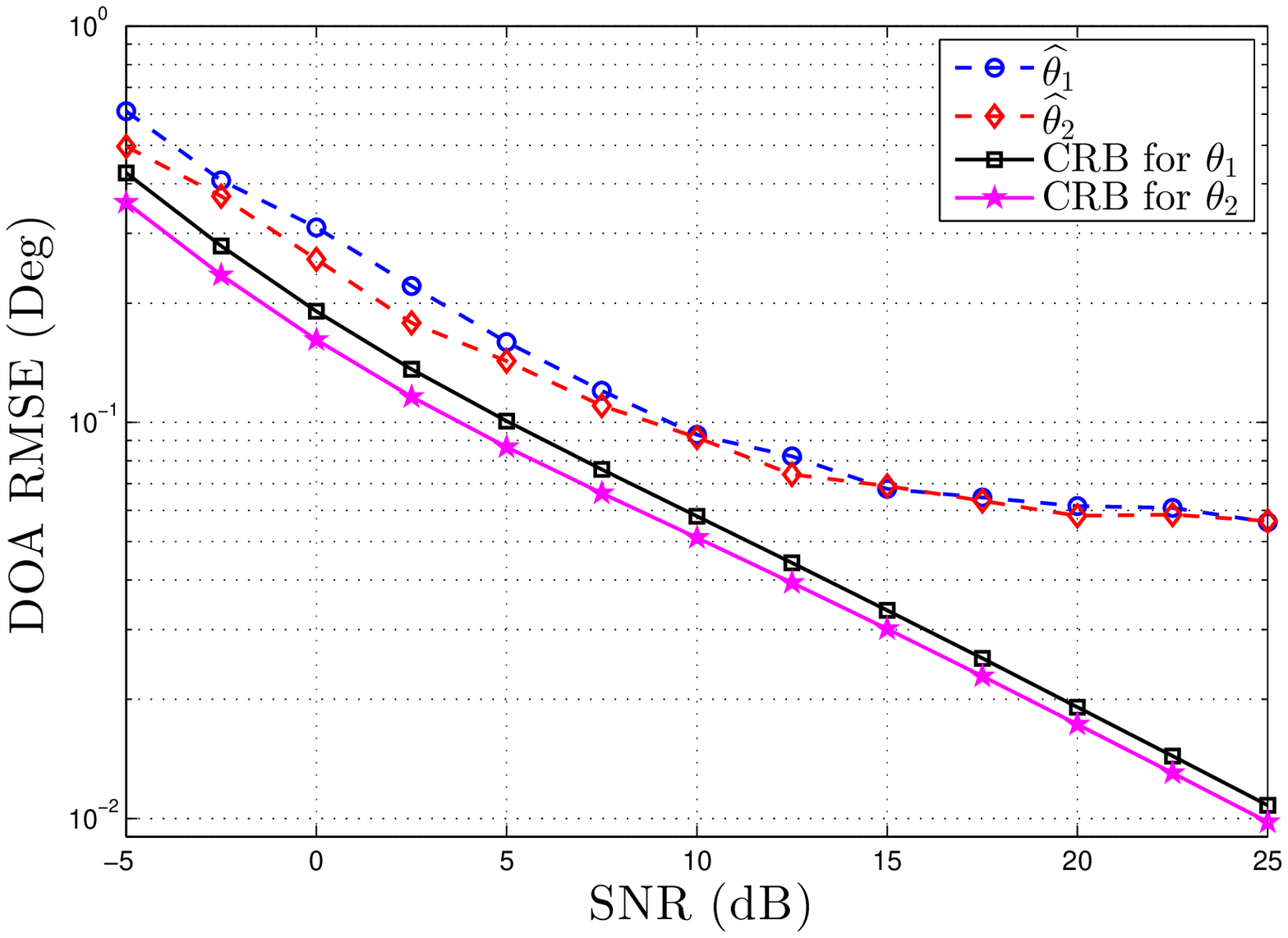}
\vspace{-0.75cm}
\caption{RMSEs for estimation of $\theta_1$ and $\theta_2$ using the proposed program \eqref{UncorrEst} as well as corresponding CRBs are plotted as a function of SNR. True $\theta_1$ and $\theta_2$ are $88.05^{\circ}$ and $91.95^{\circ}$, respectively. 500 Monte-Carlo simulations are run and $N = 500$.} \label{fig:FigSNR}
\end{figure}

In the first experiment, two uncorrelated sources at directions $\theta_1=88.05^{\circ}$ and $\theta_2=91.95^{\circ}$ impinge on the array. {\color{\RzCol}$[0^{\circ},180^{\circ}]$ is uniformly divided into 1800 points resulting in a $0.1^{\circ}$ gridding. To estimate $\theta_1$ and $\theta_2$, program \eqref{UncorrEst}, which is indeed an SOCP problem \cite{BellCW11}, is solved by CVX \cite{cvx}.}
Since \eqref{UncorrEst} is 
a square-root LASSO \cite{BellCW11}, though not optimal, based on the criterion introduced in \cite{BellCW11}, we use a fixed regularization parameter {\color{\RzCol} $\lambda_u = \frac{1}{1.1 \| \widetilde{\mathbf{s}} \|_{\infty}\sqrt{M^2 - |\Omega|}} = 0.54$, where $\widetilde{\mathbf{s}}$ denotes a fixed vector defined in \cite{BellCW11} and obtained by a simple numerical simulation \cite{BellCW11}.} 
The root mean square error (RMSE) in estimating unknown directions are reported as a function of $N$ and SNR with 500 Monte-Carlo simulations. Fig.~\ref{fig:FigN} shows the RSMEs of our approach as well as the CRBs when $N$ changes from $50$ to $10^5$ and $\TSNR$ {\color{\RzCol}is fixed to 0 dB.} As can bee seen in this figure, the proposed approach closely follows the CRB at small and medium number of samples, yet the errors remain unchanged after reaching half of the grid size. To obtain, smaller errors at larger number of measurements, one can use finer grids at the cost of an increase in computational complexity. In Fig.~\ref{fig:FigSNR}, the RMSEs and CRBs are plotted versus SNR when $N=500$. Here, we observe again a saturation in RMSEs at high SNRs which is due to the limited accuracy of the gridding.

\begin{figure}[tb]
\centering
\includegraphics[width=0.49\textwidth]{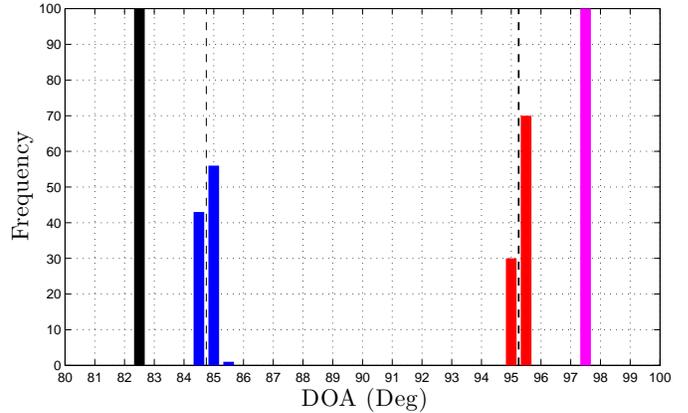}
\vspace{-0.75cm}
\caption{Histogram of the estimated directions of two near coherent sources at directions $84.75^{\circ}$ and $95.25^{\circ}$. Blue and red bars denotes the results of using programs \eqref{LEst} and \eqref{REst}, and black and magenta bars shows the results of using program \eqref{UncorrEst}. In this plot, $\TSNR = -2.5\text{ dB}$ and $N=1000$.} \label{fig:FigHist}
\end{figure}

In the second experiment, the effectiveness of programs \eqref{LEst} and \eqref{REst} in estimating the DOAs of highly correlated sources is verified. The regularization parameters $\lambda_1$ and $\lambda_2$ are numerically tuned to be 10 and 5, respectively. Two sources are at directions $\theta_1=84.75^{\circ}$ and $\theta_2=95.25^{\circ}$ with cross correlation equal to 0.99 and $\TSNR=-2.5 \text{ dB}$. Since \eqref{REst} is computationally demanding, we first use a coarse grid of $2.5^{\circ}$ and after finding two peaks from the estimated spatial spectrum, resolve \eqref{REst} with a finer grid. To be precise, let $\hat{\theta}_1^{(1)}$ and $\hat{\theta}_2^{(1)}$ denote the estimated directions with the coarse grid, in the second step, we grid the interval $[\hat{\theta}_1^{(1)} - 3^{\circ},\hat{\theta}_2^{(1)} + 3^{\circ}]$ with a fine grid of $0.5^{\circ}$. Furthermore, we also use program \eqref{UncorrEst} with a grid resolution of $0.5^{\circ}$ to estimate DOAs and show the effect of source correlation on its performance. We run 100 Monte-Carlo simulations, and the histogram of estimated DOAs for the two approaches are plotted in Fig.~\ref{fig:FigHist}. As can be seen from this plot, ignoring the correlation may cause large biases.

\vspace{-0.3mm}

\section{Conclusion} \label{sec:Con}
Based on some recent results in compressive sensing and matrix rank minimization frameworks, we proposed a DOA estimation algorithm which works well in conditions that the noise covariance matrix of the exploited array is sparse. If the emitters are uncorrelated, our approach involves solving a rather simple convex program, and we suggested an appropriate choice for the regularization parameter of this program which effectively works for any SNR and number of samples. However, when the emitters are correlated or coherent, the proposed approach leads to a computationally demanding convex optimization problem.





%

\bibliography{refs}
\bibliographystyle{IEEEbib}

\end{document}